\documentclass[preprint,endfloats]{revtex4}

\usepackage{bm}
\usepackage{pifont}
\usepackage{amsmath,amsthm,amssymb,cases}
\usepackage[dvipdfmx]{color}
\usepackage[dvipdfmx]{graphicx}

\newcommand{\bst}{(Bi$_{1-x}$Sb$_x$)$_2$Te$_3$}
\newcommand{\ef}{$E_{\rm F}$}

\begin{document}

\title{Enhanced photogalvanic current in topological insulators via Fermi energy tuning}

\author{Ken N. Okada$^{1,2}$}
\author{Naoki Ogawa$^2$}
\author{Ryutaro Yoshimi$^1$}
\author{Atsushi Tsukazaki$^{3,4}$}
\author{Kei S. Takahashi$^2$}
\author{Masashi Kawasaki$^{1,2}$}
\author{Yoshinori Tokura$^{1,2}$}

\affiliation{
$^1$Department of Applied Physics and Quantum Phase Electronics Center (QPEC), University of Tokyo, Tokyo 113-8656, Japan\\
$^2$RIKEN Center for Emergent Matter Science (CEMS), Wako, Saitama 351-0198, Japan\\
$^3$Institute for Materials Research, Tohoku University, Sendai 980-8577, Japan\\
$^4$PRESTO, Japan Science and Technology Agency (JST), Chiyoda-ku, Tokyo 102-0075, Japan.\\}

\begin{abstract}
We achieve the enhancement of circular photogalvanic effect arising from the photo-injection of spins in topological insulator thin films by tuning the Fermi level (\ef). A series of \bst\ thin films were tailored so that the Fermi energy ranges above 0.34 eV to below 0.29 eV of the Dirac point, i.e., from the bulk conduction band bottom to the valence band top through the bulk in-gap surface-Dirac cone. The circular photogalvanic current, indicating a flow of spin-polarized surface-Dirac electrons, shows a pronounced peak when the \ef\ is set near the Dirac point and is also correlated with the carrier mobility. Our observation reveals that there are substantial scatterings between the surface-Dirac and bulkstate electrons in the generation process of spin-polarized photocurrent, which can be avoided by designing the electronic structure in topological insulators.
\end{abstract}

\maketitle

Efficient spin current generation is one of the keys to materialize novel spintronics devices. In spin-split electron bands induced via spin-orbit interactions, circularly-polarized photons can drive spin-polarized charge current, or spin current, through the asymmetry in transient carrier population in the momentum space imposed by optical selection rules \cite{Meier1984, Ganichev2003}. A hallmark of this photomagnetic phenomenon, referred to as the circular photogalvanic effect (CPGE), is the reversal of current direction (and also the spin orientation) by the flip of the photon helicity, which readily finds applications in ultrafast control of spin currents. Such CPGEs have been observed in semiconductor quantum wells, e.g., GaAs/AlGaAs \cite{Ganichev2003}, and more recently in systems with stronger spin-orbit coupling, including topological insulators \cite{McIver2012,Duan2014,Kastl2015}, bulk Rashba semiconductors \cite{Ogawa2014}, and transition metal dichalcogenides \cite{Yuan2014,Eginligil2015}.

Three-dimensional topological insulators \cite{Hasan2010, Qi2011}, such as Bi$_2$Se$_3$, Bi$_2$Te$_3$ and Sb$_2$Te$_3$, host gapless Dirac electron states with spin-momentum locking at the surface \cite{Hsieh2009}. It has been demonstrated that these surface-Dirac electron states provide promising ways to correlate charge and spin degrees of freedom at high conversion efficiency \cite{Mellnik2014, Shiomi2014, Fan2014, Li2014} through the Edelstein effect \cite{Edelstein1990}, where the spins can be polarized transverse to the current direction and vice versa. The surface-Dirac electron states have an advantage over conventional Rashba-type spin-split systems, since the former have a single spin-helical Fermi circle in the momentum space, while the latter possess double Fermi circles with the opposite spin helicities. Thus the spin currents in the Rashba spin-split bands may mostly cancel, depending on the excitation condition. As mentioned above, the CPGE can further expand the versatility of this charge-spin interchange in the topological insulator by utilizing polarization of photons (or photon spins). However, some questions have been left, including what are the material parameters to maximize the photon-spin conversion efficiency. It is also important to realize these optimal parameters in chemical composition and device structure to enhance the spin current generation over the reported values in the exfoliated bulk crystal \cite{McIver2012}. In this Letter, we show the enhancement of CPGE currents in \bst\ (BST) thin films by precisely tuning the Fermi energy (\ef) across the Dirac cone at room temperature in ambient atmosphere. The origin for this enhancement in the spin current generation is discussed in terms of the scattering between the surface and bulk states upon photoexcitation.

The BST thin films were fabricated on insulating InP (111) substrates (band bap 1.34 eV) by using molecular beam epitaxy \cite{Yoshimi2014}, immediately followed by {\it ex-situ} atomic-layer-deposition of 10-nm-thick AlO$_x$ layers to prevent sample degradation. The composition dependence of the CPGE was evaluated in the films with the fixed thickness of 8 nm, where the bulk volume is minimized and the top and bottom surface states remain decoupled \cite{Yoshimi2015, Zhang2010}. In the photocurrent measurement a cw laser (1.17 eV), chopped at 397 Hz, was irradiated on the films with the photon polarization controlled by a quarter-wave plate, as illustrated in Fig. 1(a) (see also Fig. S1 in Supplemental Material \cite{supple}). The laser spot with the diameter around 1 mm was located between the end electrodes separated by $\sim$ 3 mm, to avoid extrinsic photocurrents \cite{McIver2012}. The photocurrent in the InP substrates was negligibly small at the photon energy used. The short-circuit zero-bias photocurrent was measured by a lock-in amplifier, with the absolute phase deduced by using a reference photo diode to determine the current direction. All the photocurrent measurements were performed at room temperature.

Figures 1(b)-1(d) illustrate the representative CPGE driven by surface Dirac electrons observed in the BST ($x$ = 0.80) film, which has \ef\ in the bulk band gap, i.e., on the Dirac cone just above the Dirac point (DP), as will be shown later. The emergence of the CPGE is indicated by the distinct difference between the photocurrent induced by the photons with the opposite helicity [Fig. 1(b)]. The measured photocurrent can be well expressed by the formula demonstrated to be useful by the previous report on Bi$_2$Se$_3$ thin flakes \cite{McIver2012};
\begin{eqnarray}
J=C{\rm sin}\ 2\alpha +L_1{\rm sin}\ 4\alpha +L_2{\rm cos}\ 4\alpha +D
\end{eqnarray}
,where $\alpha$ is the angle of the quarter-wave plate, $C$ the amplitude of the CPGE current, $L_1$ the linear photogalvanic currents \cite{Olbrich2014}, $L_2$ photocurrent modulations related to Fresnel factors, $D$ thermal and photon drag effects \cite{McIver2012}. The CPGE current ($C$) shows a cosine-like function of the azimuthal angle $\varphi$ at the incident angle $\theta$ = 45 deg. [Fig. 1(c)] and vanishes at the normal incidence ($\theta$ = 0 deg.) [see also Fig. 4(a)], consistent with the helical spin texture of the Dirac cone \cite{Hsieh2009}. The CPGE current varies linearly with the light intensity [Fig. 1(d)].

To understand how the CPGE current changes by tuning the position of \ef\, we compare the photocurrent in the BST thin films with the Sb content $x$ varied from 0.0 to 1.0; the results are summarized in Fig. 2. The transport properties shown in Figs. 2(a) and 2(b) indicate that \ef\ is systematically modified across the DP. We note that three samples ($x$ = 0.82, 0.86 and 0.88, represented with broken lines) were fabricated in a slightly different condition during additional experimental runs, and make some deviations from the overall trends in $R_{xx}$ [Fig. 2(a)] and mobility [Fig. 3(a)]. However, these fluctuations will not affect our conclusions (see below). For Bi$_2$Te$_3$ ($x$ = 0.0), since $R_{xx}$ shows metallic behavior and $R_{yx}$ is negative, \ef\ lies in the conduction band. With increasing $x$ from 0.0 to 0.84, $R_{xx}$ and the absolute value of the Hall coefficient ($|R_{\rm H}| = |R_{yx}|/B$) increases, which means the shift of \ef\ towards the DP. As $x$ increases from 0.84 to 0.86, $R_{yx}$ switches from negative to positive, signifying the change of the carrier type from electron to hole with \ef\ traversing across the DP. By extracting the position of \ef\ from the carrier density ($n = 1/e|R_{\rm H}|$) at 2 K [Fig. 3(a)] (Supplemental Material \cite{supple}), we conclude that \ef\ lies within the bulk band gap from $x$ = 0.76 to 0.88. Note that the samples with $x$ = 0.84 to 0.88 show the smaller $R_{yx}$ [Fig. 2(b)] and seemingly have larger carrier densities [Fig. 3(a)] than those in the adjacent samples ($x$ = 0.82 and 0.90). However, this is perhaps due to the coexistence of electron and hole puddles when \ef\ is close to the DP \cite{Yoshimi2015}. In Sb$_2$Te$_3$ ($x$ = 1.0), a metallic behavior is recovered in $R_{xx}$ and therefore \ef\ is located in the bulk valence band. Such a systematic variation of the transport properties ensures the precise tuning of \ef\ with the Sb content $x$ \cite{Zhang2011}. The observed photocurrents with varying light polarization ($\alpha$) are plotted in Fig. 2(c) for all the samples to see the impact of \ef\ on the spin current generation. Note that the polarization-independent term [$D$ in equation (1)] is subtracted here. It is clearly seen that the photocurrent is dramatically enhanced when \ef\ locates within the band gap. Figure 3 summarizes the variation of the CPGE current [$C$ in Eq. (1)] with $x$, in comparison with the carrier mobility ($\mu = |R_{\rm H}|/R_{xx}$) and density. As readily seen in the raw data in Fig. 2(c), the CPGE current is largely enhanced, forming a pronounced peak as a function of $x$, when \ef\ is located within the bulk band gap. Note that the CPGE current rapidly decreases in the both sides of the peak [Fig. 3(a)], which is not apparent in Fig. 2(c) due to the $L_1$ and $L_2$ components in Eq. (1).

From the distinct \ef\ dependence of the CPGE [Fig. 3(a)], we assume that the CPGE current is carried by the electrons/holes photoexcited near the Fermi level. The microscopic process of the CPGE enhancement can be attributed to the suppression of the scattering of the surface Dirac electrons to the bulk channel [Figs. 3(b) and 3(c)]. As \ef\ is shifted lower in going from $x$ = 0.60 to 0.70, the mobility increases [Fig. 3(a)] since the surface Dirac electrons begin to dominate over the bulk electrons. However, since the \ef\ still lies in the conduction band, spin-polarized surface-Dirac electrons excited at the Fermi level are easily scattered into the spin-degenerate bulk conduction band by phonons and impurities \cite{Reijnders2014, Zhu2012, Chen2013, Wang2012}, leading to the minimal CPGE current [Fig. 3(b)]. When \ef\ is within the bulk band gap (from $x$ = 0.76 to 0.88), the CPGE current unveils as the scattering to the bulk channel is suppressed [Fig. 3(c)]. Finally, when \ef\ is lowered into the bulk valence band for $x$ = 0.90 and 1.0, the CPGE current is again suppressed due to the scattering into the bulk states.

It should be noted that the CPGE current also exists at $x$ = 0.0 (Bi$_2$Te$_3$). We speculate that this comes from much larger bulk-carrier absorption in this end compound than the other compositions (see Fig. S2 in Supplemental Material \cite{supple}), which may reduce the cancellation between the CPGE current generated at the top and bottom surfaces (see the discussions on Fig. 4 below).

The maximal yield of the CPGE current observed above ($\sim$ 10 pAW$^{-1}$cm$^2$), which is calculated with the spot size taken into account, is comparable to those in semiconductor quantum wells (1 $\sim$ 100 pAW$^{-1}$cm$^2$) \cite{Ganichev2003,Ganichev2001,Ganichev2002}, and several times larger than the reported value in Bi$_2$Se$_3$ thin flakes \cite{McIver2012}. Furthermore, by normalizing the CPGE current by the number of photons absorbed, we estimate the generation efficiency for our BST film with the \ef\ near the DP to be twenty times as large as that for the Bi$_2$Se$_3$ flake \cite{McIver2012}. We note that the sign of the CPGE current remains unchanged even when the \ef\ position changes between above and below the DP [Fig. 3(a)], although the spin helicity itself changes \cite{footnote}. This is quite reasonable because (i) the direction of the photocurrent is determined by the group velocity, and (ii) the group velocity for electrons with a specific spin orientation has the same sign, irrespective of their energy above or below the DP. Therefore, carriers with the specific spin state generated by circularly-polarized photons should flow in the same direction. This has been reported in other spin-to-charge conversion experiments, such as current-induced spin polarization \cite{Lee2015, Kondou2015}.

Finally, we discuss the contributions of the top and bottom surface states to the observed CPGE current. For the film thickness used (8 nm), the observed CPGE current should be the sum of the currents generated at the both surfaces, which flow in the opposite directions due to the opposite spin helicities of the Dirac cone. The CPGE current would be detectable solely by the difference in the photon number at both surfaces due to the finite absorption in between, if both the surface states were physically equivalent. To check this cancellation, we measured the incident angle dependence of the CPGE current for the front [Fig. 4(a)] and the back surface illumination [Fig. 4(b)] for the film with $x$ = 0.84 (see Supplemental Material for detail \cite{supple}). For the back illumination, we define the incident angle $\psi$, whose positive direction is opposite to that of $\theta$ [Figs. 4(a) and 4(b)]. If the both surface states were equivalent, the CPGE current for the back illumination would have the opposite sign to that for the front illumination. However, in reality, the back illumination [Fig. 4(b)] revealed the same sign of the signal, suggesting that one of the surface states can generate CPGE current more efficiently than the other. The same trend is observed in other compositions (see Supplemental Material \cite{supple}). We speculate that such inequivalence between the top and bottom surface states can be ascribed to the difference in scattering time between them due to the quality of the top and bottom layers depending on growth process. 

In summary, we have achieved the optimization of the circular photogalvanic current in topological insulator thin films by tuning the Fermi energy near the Dirac point. Such an enhancement of the spin photocurrent is ascribed to the suppression of the scattering of the surface Dirac electrons to the bulk channel. The present study proves that topological insulators provide a potential platform for novel opto-spintronic devices.
\\
\\
The authors are grateful to Y. Takahashi and M. S. Bahramy for stimulating discussions. This research was supported by the Japan Society for the Promotion of Science through the Funding Program for World-Leading Innovative R \& D on Science and Technology (FIRST Program) on 'Quantum Science on Strong Correlation' initiated by the Council for Science and Technology Policy and by JSPS Grant-in-Aid for Scientific Research(S) No. 24224009 and 24226002. K.O. is supported by RIKEN Junior Research Associate Program.

\clearpage

\begin{figure}[t]
\begin{center}
\includegraphics[width=120mm]{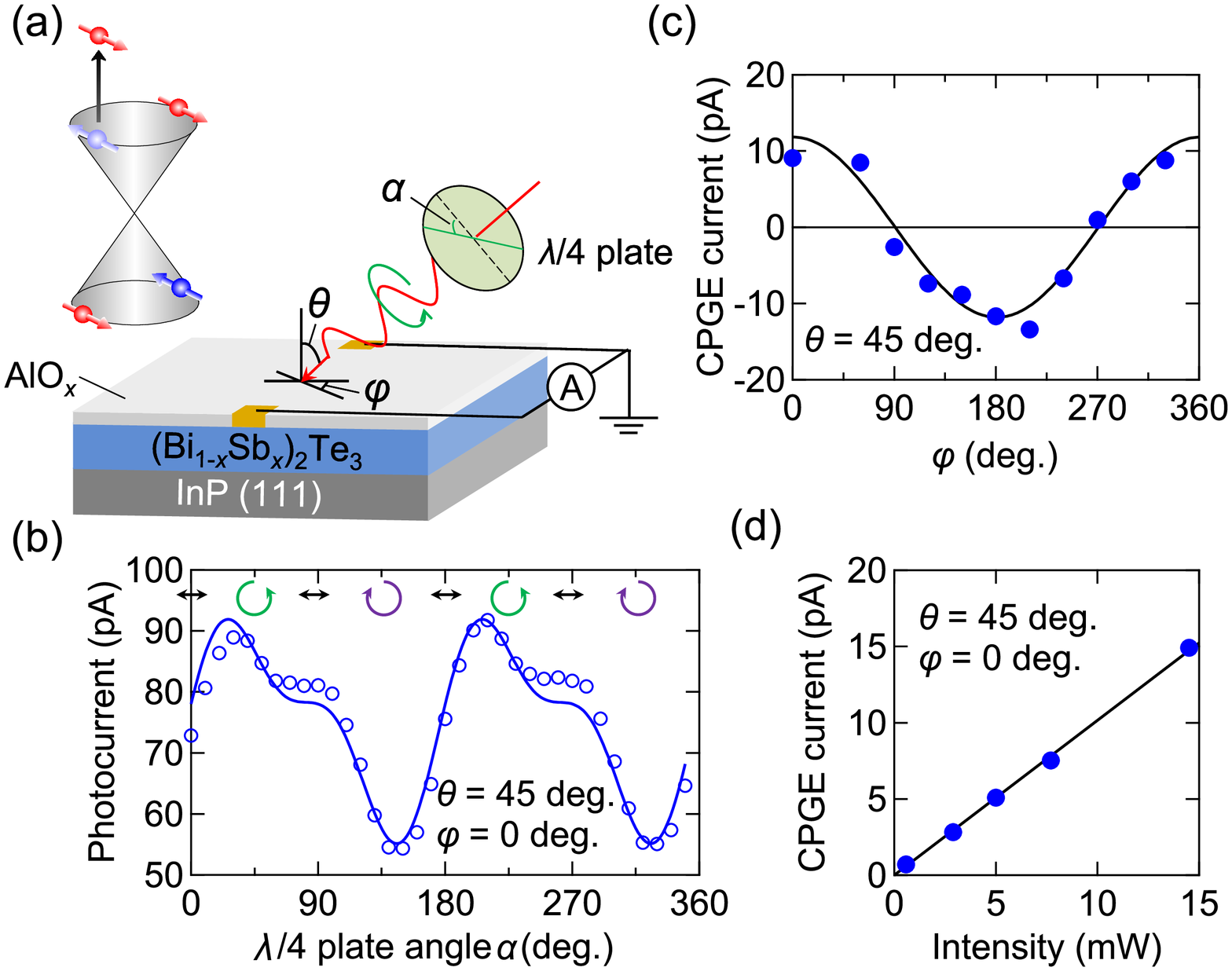}
\end{center}
\caption{(Color online) (a) Schematic illustration of the experimental setup. The laser is irradiated on the sample at the incident angle $\theta$ and the azimuthal angle $\varphi$ . (b)-(d) Circular photogalvanic effect (CPGE) driven by the surface Dirac electrons in the \bst\ (BST, $x$ = 0.80) film. (b) Photon polarization dependence at $\theta$ = 45 deg. and $\varphi$ = 0 deg. The solid line is a fit to Eq. (1) in the text. (c) Azimuthal angle dependence at  $\theta$ = 45 deg. The solid line represents cosine dependence. (d) Laser intensity dependence.}
\label{fig:fig1}
\end{figure}

\clearpage

\begin{figure}[t]
\begin{center}
\includegraphics[width=120mm]{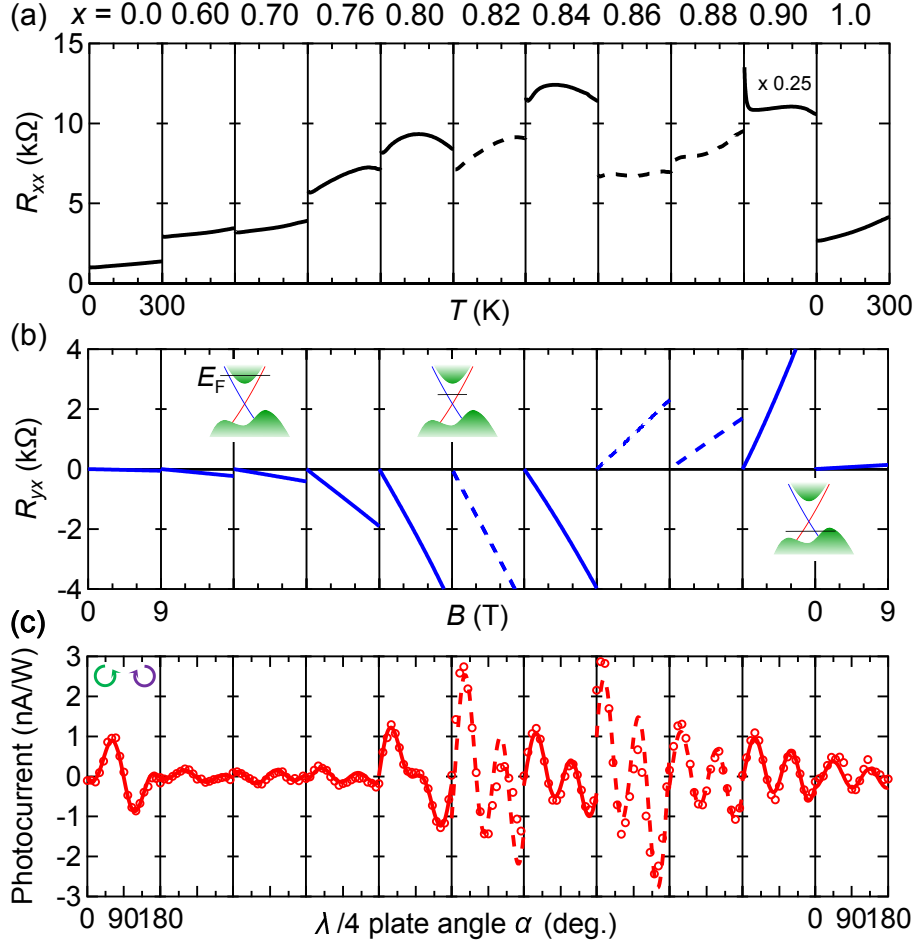}
\end{center}
\caption{(Color online) Evolution of the transport and CPGE characteristics by tuning the Fermi energy in BST thin films. (a) Temperature dependence of the longitudinal resistance $R_{xx}$. (b) Field dependence of the Hall resistance $R_{yx}$ at 2 K. (c) Polarization dependent photocurrent measured at $\theta$ = 45 deg. and $\varphi$ = 0 deg. The solid lines are fit to Eq. (1). The samples with the data represented with broken lines were fabricated in a different chamber condition.}
\label{fig:fig2}
\end{figure}

\clearpage

\begin{figure}[t]
\begin{center}
\includegraphics[width=120mm]{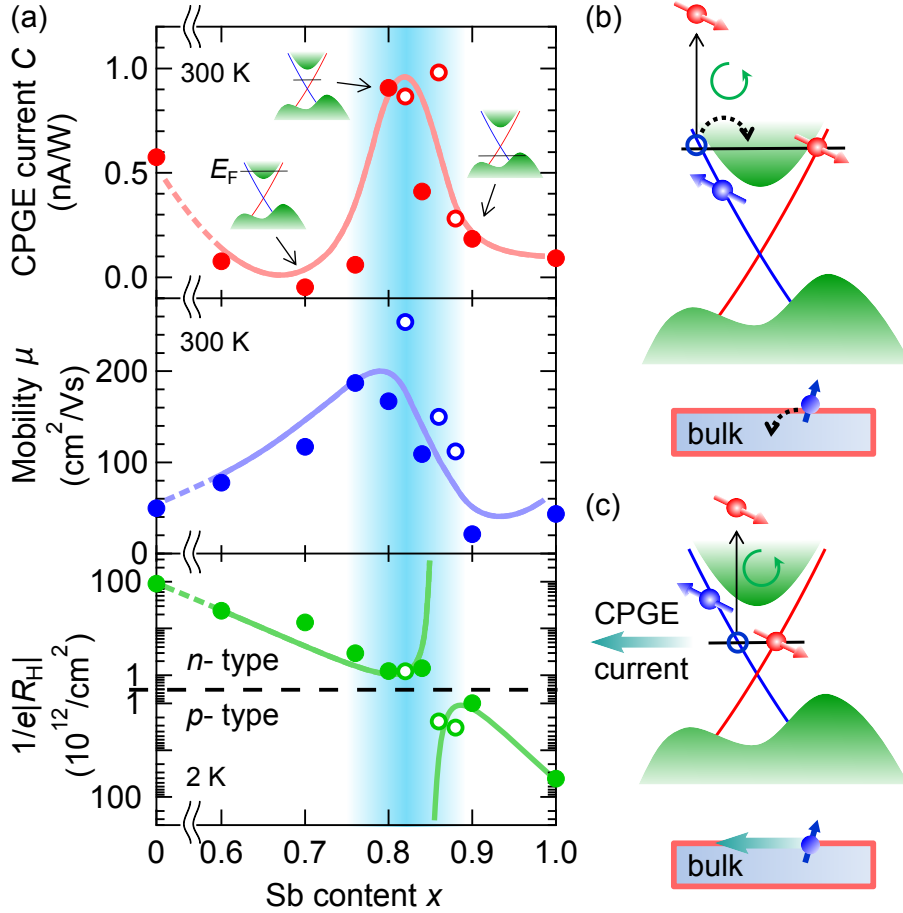}
\end{center}
\caption{(Color online) (a) The CPGE current as a function of the Sb content $x$, plotted together with the mobility and carrier density. The carrier density is measured at 2 K to precisely estimate the Fermi energy. The samples with the data represented with open circles were fabricated in a different chamber condition. The solid lines are guides to the eyes. The shaded area indicates the compositions where the Fermi energy is in the bulk band gap. (b), (c) Schematic illustrations for the CPGE process when the Fermi energy locates in the bulk conduction band (b) and in the bulk band gap (c).}
\label{fig:fig3}
\end{figure}

\clearpage
\begin{figure}[t]
\begin{center}
\includegraphics[width=120mm]{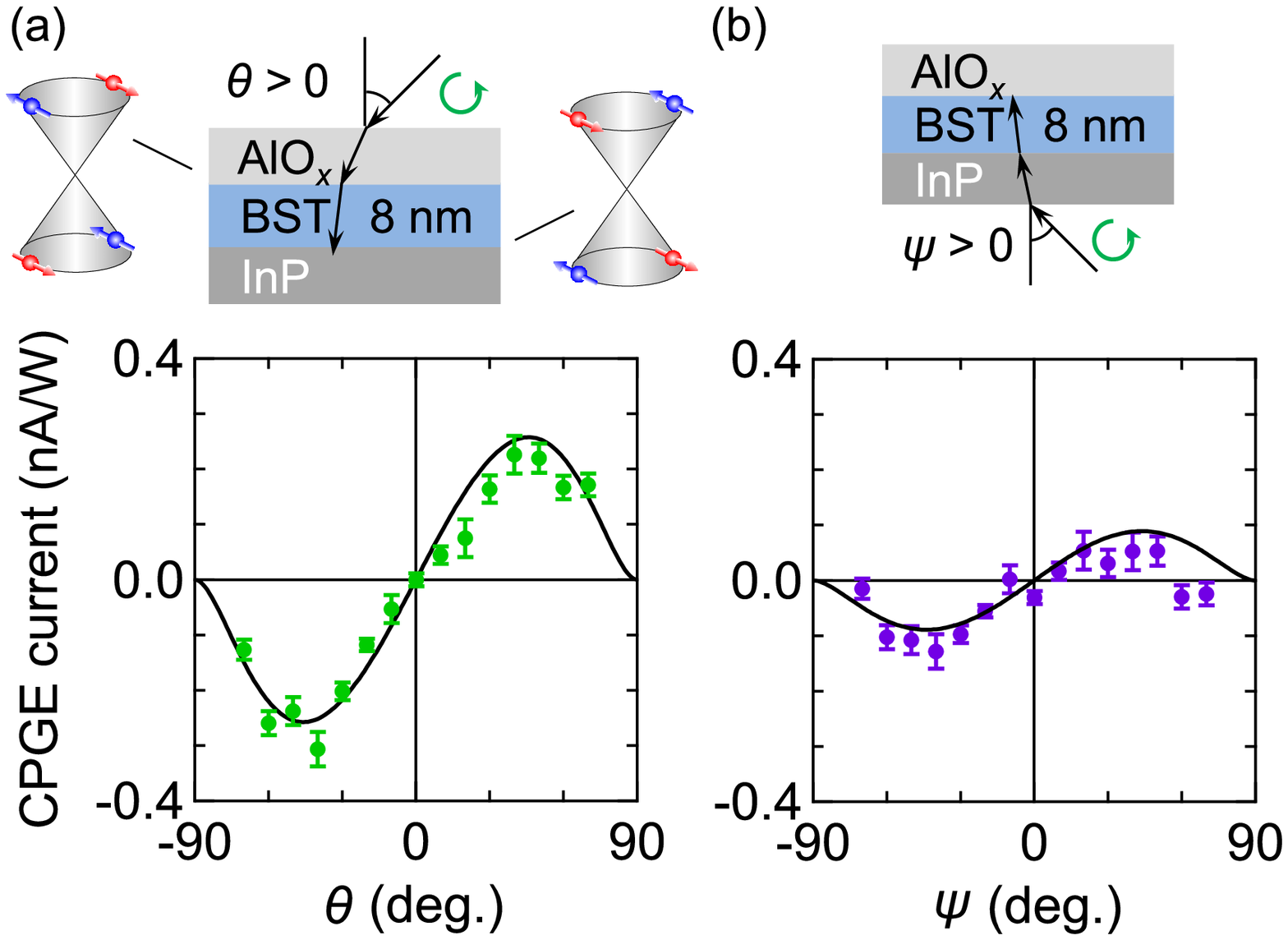}
\end{center}
\caption{(Color online) Incident angle dependence of the CPGE current for (a) front or (b) back illumination. The solid lines represent the theoretical fit (See Supplemental Material \cite{supple}).}
\label{fig:fig4}
\end{figure}


\begin{thebibliography}
\bibitem{footnote}The CPGE current for $x$ = 0.70 [Figs. 2(c) and 3(a)] shows a small negative value. However, it is difficult to quantitatively discuss this point because its amplitude is close to our experimental resolution, and the \ef\ is located in the bulk conduction band.
\end{thebibliography}
\end{document}